\documentclass[twocolumn]{svjour3}       
\smartqed  
\usepackage{graphicx}
\usepackage{cite}
\usepackage{amsmath,amssymb,amsfonts}
\usepackage{algorithmic}
\usepackage{textcomp}
\usepackage{xcolor}

\begin{document}

\title{A Review of Machine Learning Classification Using Quantum Annealing for Real-world Applications\\}


\author{Rajdeep Kumar Nath         \and
        Himanshu Thapliyal* \and Travis S. Humble 
}


\institute{Rajdeep Kumar Nath \at
              University of Kentucky \\
              Lexington, KY, USA\\         
               \and
               Himanshu Thapliyal* \at
              University of Kentucky \\
              Lexington, KY, USA\\
              Corresponding Author: hthapliyal@uky.edu
           \and
           Travis S. Humble \at
              Oak Ridge National Laboratory\\
			Oak Ridge, TN, USA \\
             }

\date{Received: date / Accepted: date}

\maketitle

\begin{abstract}
Optimizing the training of a machine learning pipeline helps in reducing training costs and improving model performance. One such optimizing strategy is quantum annealing, which is an emerging computing paradigm that has shown potential in optimizing the training of a machine learning model. The implementation of a physical quantum annealer has been realized by D-Wave systems and is available to the research community for experiments. Recent experimental results on a variety of machine learning applications using quantum annealing have shown interesting results where the performance of classical machine learning techniques is limited by limited training data and high dimensional features. This article explores the application of D-Wave’s quantum annealer for optimizing machine learning pipelines for real-world classification problems. We review the application domains on which a physical quantum annealer has been used to train machine learning classifiers. We discuss and analyze the experiments performed on the D-Wave quantum annealer for applications such as image recognition, remote sensing imagery, computational biology, and particle physics. We discuss the possible advantages and the problems for which quantum annealing is likely to be advantageous over classical computation.

\end{abstract}
\keywords{Classification \and Machine Learning \and Optimization \and  Quantum Annealing  \and Quantum Computing}


\section{Introduction}
Machine learning techniques to explore and harness the power of data have found application in health care, finance, autonomous driving, security, etc., and are continuing to make their impact deeper every single day \cite{intro_002}\cite{intro_003}\cite{intro_004}. However, this technological boom is accompanied by unprecedented challenges. These challenges arise from several factors such as the scale of data being generated, hardware limitations, computational complexity, and cost \cite{intro_001}. Although recent advances in hardware technology have increased computational capability significantly, this is no match for the amount of data generated globally, which is rapidly increasing, as well as the amount of stored data, which is increasing at the rate of about 20\% per year \cite{001}. The current trend of technological advancement in terms of computational power will become saturated in dealing with the massive scale of data which will result in increased cost, and more importantly, the maximum utilization of data won't be possible in the future \cite{intro_001}. To deal with this challenge, quantum computing is seen as a promising alternative that can boost the computational prowess in utilizing the power of data to the full extent \cite{intro_005}.

\begin{figure}[h]
\centering
\includegraphics[trim=1cm 0cm 0cm 0cm, scale=0.25]{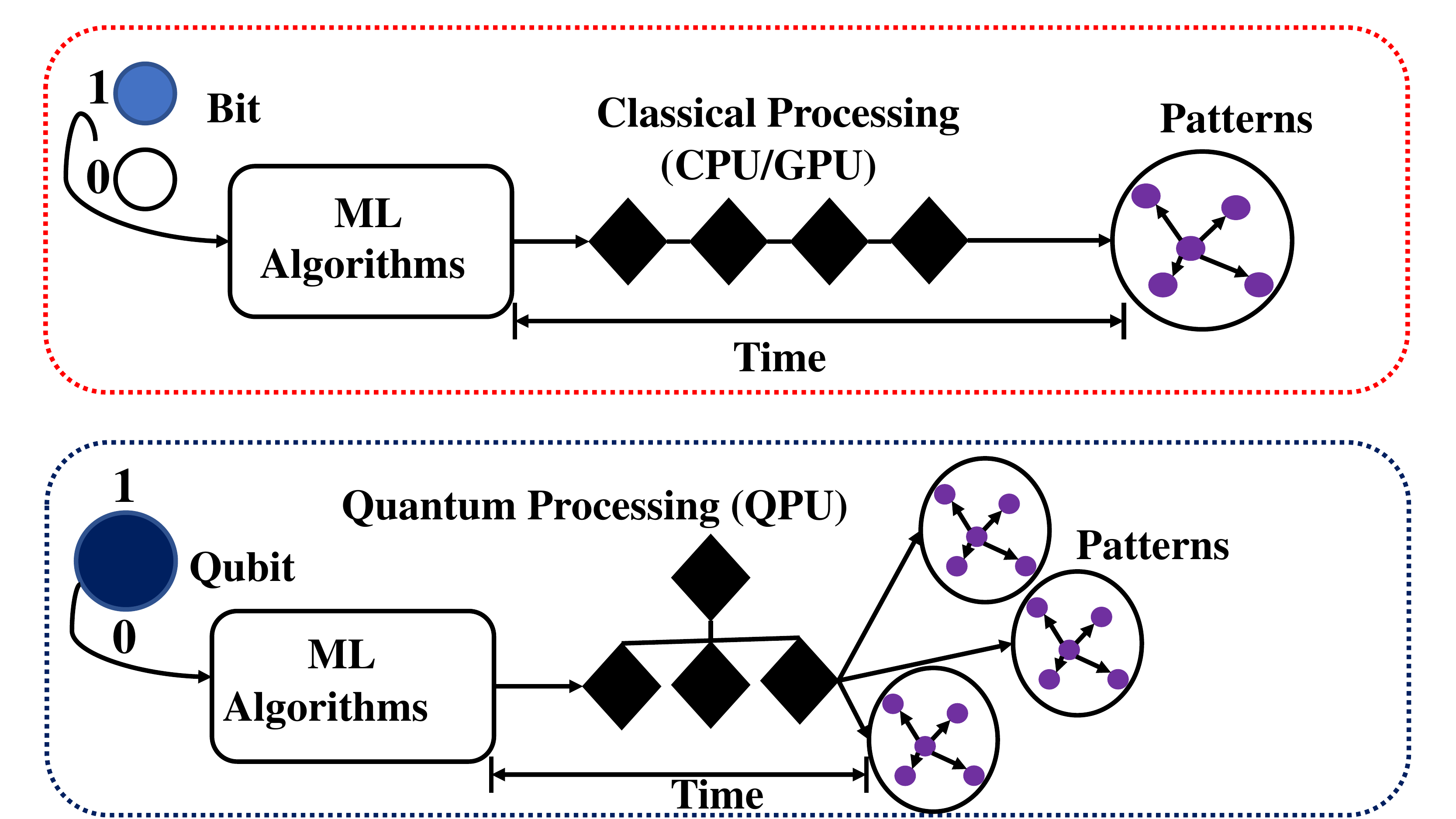}
\caption{Processing in classical (top) and quantum systems (bottom). QPU can generate many-solutions and counter-intuitive solutions faster than classical processing units such as CPU/GPU.}
\label{fig1}
\end{figure} 

Most of the real-world problems that machine learning seeks to solve are non-convex in nature. These types of problems are hard to converge to an optimal solution using classical algorithms because of the possibility of the presence of several local minima and saddle points \cite{intro_006}. This computational challenge limits the ability of machine learning models to train efficiently especially in the presence of a large number of variables. To deal with this problem, researchers have been exploring alternative computing techniques such as quantum computing to augment the power of machine learning algorithms \cite{003}\cite{015}\cite{017}. Quantum computing is inherently suited to carry out complex computation which is not possible using classical computation \cite{034}. Quantum computing processes information in the form of qubits (quantum bits). This is different from a classical bit in the sense that a classical bit can take the value of either 0 or 1. However, a qubit can be represented as a combination of two values at the same time. This principle is called superposition which allows quantum computation to achieve exponential speedup over its classical counterpart \cite{002}. Another property of quantum computation that establishes superiority over classical computation is the phenomenon of quantum tunneling. Because of quantum tunneling, a quantum system can navigate through solution space more efficiently by overcoming long and thin energy barriers \cite{004}. These characteristics of quantum computing have led to the hypothesis that QPU (Quantum Processing Unit) can be used to discover more interesting and counter-intuitive patterns for machine learning classification task than that obtained using classical processing units such as CPU (Central Processing Unit) or GPU (General Processing Unit) \cite{005}(Figure \ref{fig1}).

The main contributions of this article are as follows:

\begin{itemize}
\item The integration of D-wave's quantum annealer as an optimization subroutine has been analyzed and discussed in the context of machine learning classification tasks.
\item Application domain where quantum annealing is applied for real-world classification tasks has been identified and analyzed from existing literature.
\item Finally, the possible advantages of using quantum annealer as an optimization subroutine for machine learning classification have been analyzed based on the findings from recent research.
\end{itemize}

Developing a hybrid classical-quantum system for machine learning classification tasks is expected to perform better than using a classical machine learning pipeline under data uncertainty conditions. Some of these conditions are limited availability of training data, modeling under high dimensional feature variables, better generalization to unknown data, and reduced training time and cost. This article explores the effectiveness and usefulness of quantum annealing as an optimization routine in machine learning pipelines for classification tasks. We present a comprehensive survey on quantum annealing for real-world machine learning classification tasks based on recent research works presented in the literature.

The paper is organized as follows: Section \ref{background_qa} discusses the theoretical background of quantum annealing and also the realization of the quantum annealing in D-Wave's quantum computer. Section \ref{dwave_qpu} discusses the hybrid classical-quantum computing system using D-Wave's quantum computer. This section details the D-Wave's implementation and translation of real-world problem instances into the quantum processing unit of D-Wave. Section \ref{works_qa} will explore the existing literature in the area that has used quantum annealing for machine learning classification tasks. Section \ref{advantage_qa} discusses the possible advantages of using quantum annealing over classical techniques. Section \ref{conc} concludes the article.

\section{Background on Quantum Annealing}
\label{background_qa}
\begin{figure}[h]
\centering
\includegraphics[trim=1cm 0cm 0cm 0cm, scale=0.28]{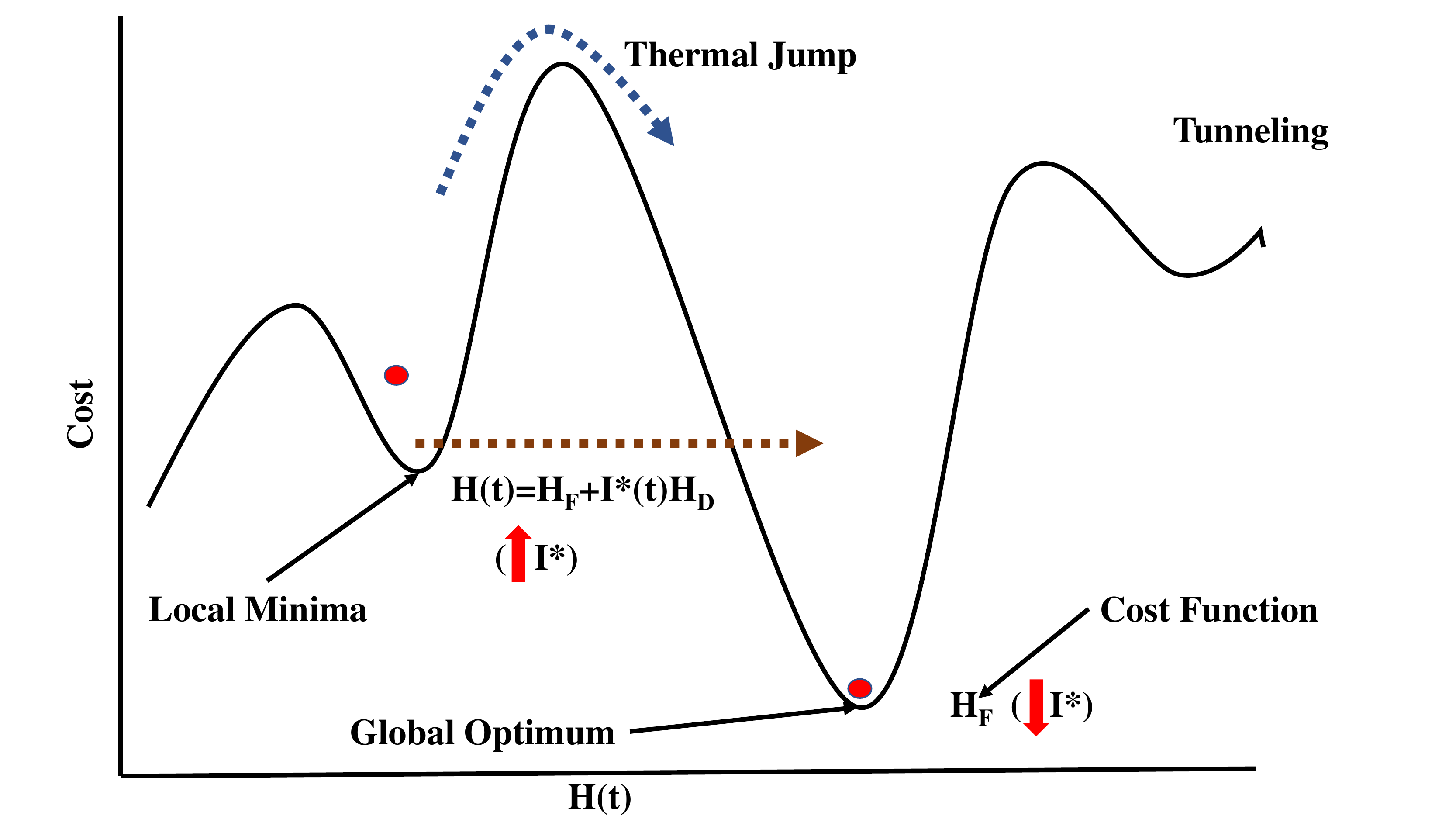}
\caption{Solution landscape of a cost function represented by a time-dependent Hamiltonian. High values of the transverse field coefficient drives tunneling phenomenon in order to escape local minima and move towards the optimal solution.}
\label{qa_background_label}
\end{figure}

Quantum annealing is a heuristic search algorithm for finding the lowest energy state by traversing over the solution landscape \cite{035}. The lowest energy state is achieved through the evolution of a time-dependent Hamiltonian. Quantum annealing uses quantum tunneling to traverse through the solution space more efficiently than classical annealing to reach the optimal state \cite{007,008}. As visualized in Figure \ref{qa_background_label}, the time-dependent Hamiltonian can be represented as a landscape of the cost of the solution. The two components of the time-dependent Hamiltonian is the final Hamiltonian ($H_{F}$), that is the ground state or the optimal solution to the problem and the second component is a transverse field Hamiltonian ($H_{D}$) which is scaled by a time-dependent coefficient also called transverse field which is initialized at a high value and then decreased to zero. Large values of the transverse field allow the algorithm to avoid local minima and tunnel through large energy barriers to move towards the ground state. This same phenomenon is achieved in classical annealing by thermal jumps. It has been theoretically proven that quantum annealing is guaranteed to converge to the optimal solution and also the convergence rate of quantum annealing was faster than that of classical annealing \cite{009}.

Even though quantum annealing optimization algorithm has proven to be theoretically superior to classical annealing, a classical implementation of such an algorithm is both costly and inefficient \cite{007}. Whereas some operations are more efficient and faster using quantum computing, there are also several operations for which classical systems are inherently better than quantum systems. A hybrid system that uses both classical and quantum computing is gaining immense popularity in recent years. In the context of machine learning, training becomes extremely expensive and inefficient as the dimension of the feature vector increases which in turn makes it difficult to find meaningful patterns. These types of computationally challenging tasks could be offloaded to quantum processors for optimization \cite{036}. Recent advances made in the design of physical quantum annealer have made it possible to explore the practical application of quantum algorithms. D-Wave system is currently leading the market of commercial quantum computers which uses quantum annealing for computation. We will now present a brief overview of the implementation of quantum annealing on D-Wave systems.

\subsection{Quantum Annealing in D-Wave Systems}
\begin{figure}[h]
\centering
\includegraphics[trim=1cm 0cm 0cm 0cm, scale=0.25]{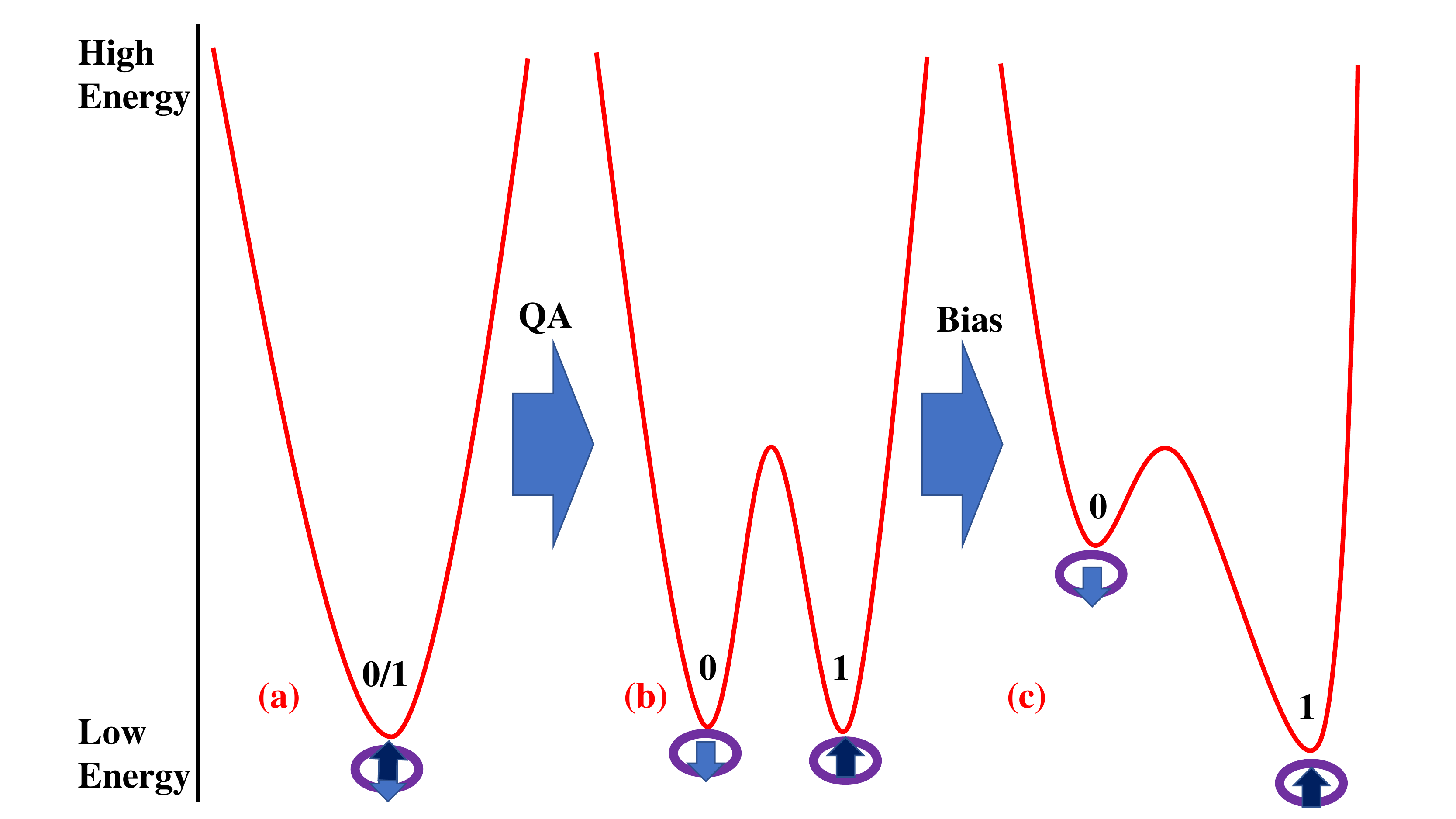}
\caption{Energy diagram of the quantum annealing process of a single qubit. }
\label{qa_physics_label}
\end{figure}

In D-Wave's implementation of quantum annealing, a qubit initially remains in a superposition state. At the end of annealing, the qubit goes from the superposition state to either 0 state or 1 state. Figure \ref{qa_physics_label} shows the energy diagram depicting the physics associated with the process of quantum annealing and how the qubits attain the lowest energy state. 

\begin{figure*}[h]
\centering
\includegraphics[trim=1cm 0cm 0cm 0cm, scale=0.4]{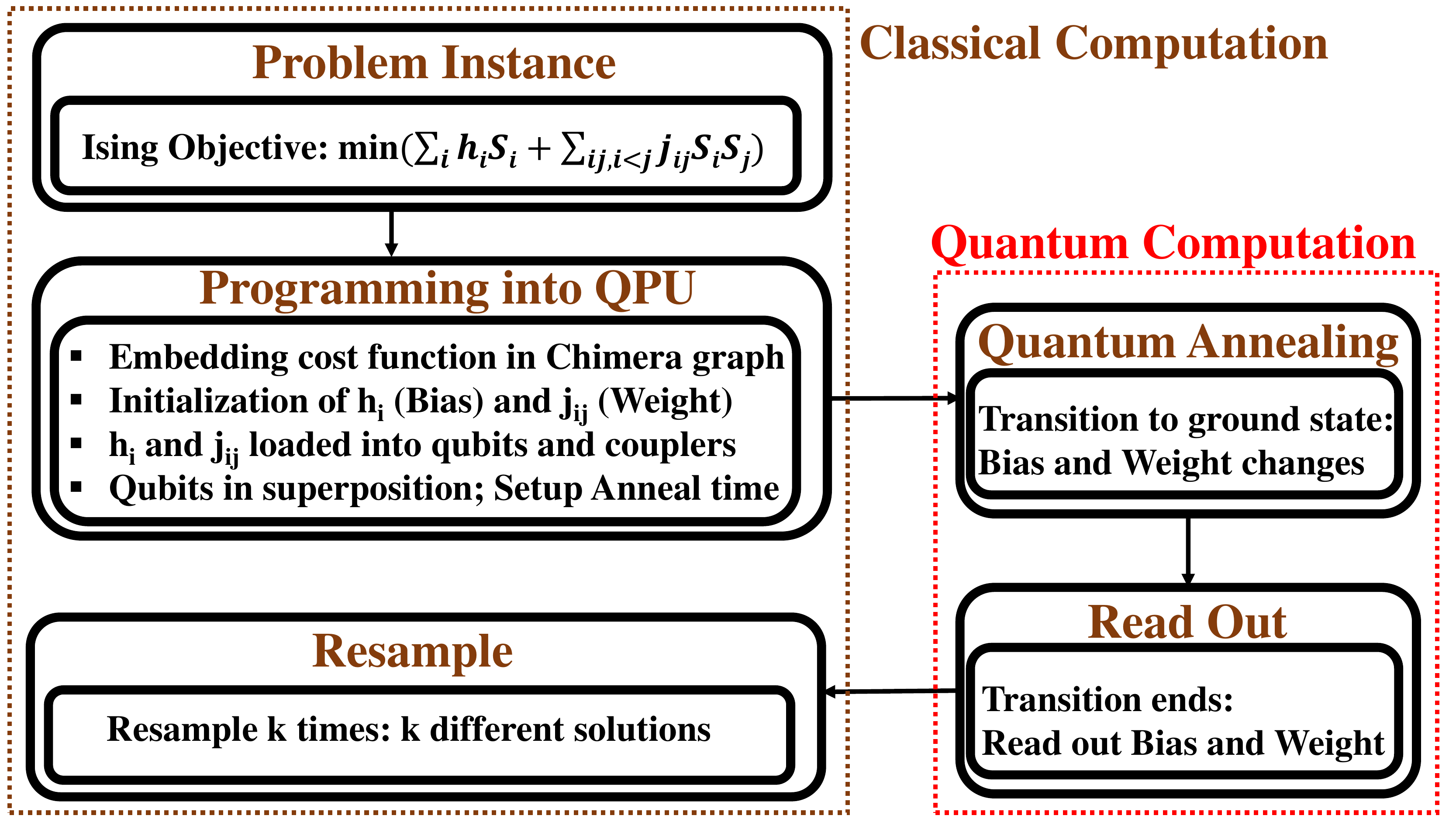}
\caption{Overview of the steps required for computation using D-Wave's QPU \cite{007}}
\label{qpu_overview_label}
\end{figure*} 
Figure \ref{qa_physics_label} shows three configurations of the energy diagram. The initial configuration (a) consists of only one valley, with the qubit in the superpositioned state. When quantum annealing is run, a barrier is raised which gives rise to the formation of a double-well potential (configuration (b)). In this configuration, the qubits have an equal probability of ending up either in the 0 state (low point of the left valley) or in 1 state (low point of the right valley). An interesting feature of the quantum annealing processor is that the probability with which a particular qubit will fall either in 0 state or 1 state can be controlled by applying an external magnetic field to the qubits also called the bias. In other words, the qubits can minimize their energy under the influence of the external magnetic field or bias. 

Coupling is another important feature of the quantum annealing processor. Coupling is the method through which two coupled qubits can be made to be either in the same state or in a different state, that is both 0 or 1 or one 0 and the other 1 or vice versa. This phenomenon of coupling is known as entanglement in quantum computing. For example, when two qubits are entangled, they are considered as one object but with four possible states or combinations. Hence, a two-qubit system will have potential with four states where each state represents a different combination. This defines the energy landscape of the two qubits governed by the relative energy between the qubits. The relative energy between the qubits depends on the biases of each qubit and the coupling between them. The programmer chooses the bias and coupling to encode a specific problem instance. 

\section{Quantum Processing Unit of D-Wave Systems}
\label{dwave_qpu}
In the previous section, we discussed the basics of computation using quantum annealing and discussed some key concepts such as qubits, bias, and coupler. These concepts are necessary for understanding the architecture of the D-Wave's Quantum Processing Unit (QPU) and the application of D-Wave's QPU in solving real-world problem instances. Figure \ref{qpu_overview_label} shows an overview of the steps involved in performing computation using D-Wave's QPU. The overall process can be viewed as a hybrid system consisting of both classical computation and quantum computation. 

\subsection{Classical Computation}
The classical computation consists of three main parts, (i) Problem initialization, (ii) Programming the problem instance into the quantum annealing hardware through a software interface, and (iii) Resampling. The first two steps take place before annealing and the last step takes place post-annealing. 
\subsubsection{Problem Instance}
To solve a problem with QPU, the problem needs to be translated into either an Ising objective function or a QUBO objective function. Translation between these two types of functions is straightforward. Both these functions support only quadratic functions, the reason for which will be clear in section \ref{soft_interface}. For example, considering an Ising objective seeks to find real-valued weights $h_{i}$ and $J_{ij}$ to n-spin variables that will minimize the objective function. In the context of machine learning classification, $h_{i}$ and $J_{ij}$ can be thought of as biases and weights and the n-spin variables can be thought of as the features or variables that are used for classification. After the problem instance is defined, it is programmed into the quantum annealing hardware using a software interface. 
\subsubsection{Programming into QPU}
\label{soft_interface}
\begin{figure*}[h]
\centering
\includegraphics[trim=1cm 0cm 0cm 0cm, scale=0.5]{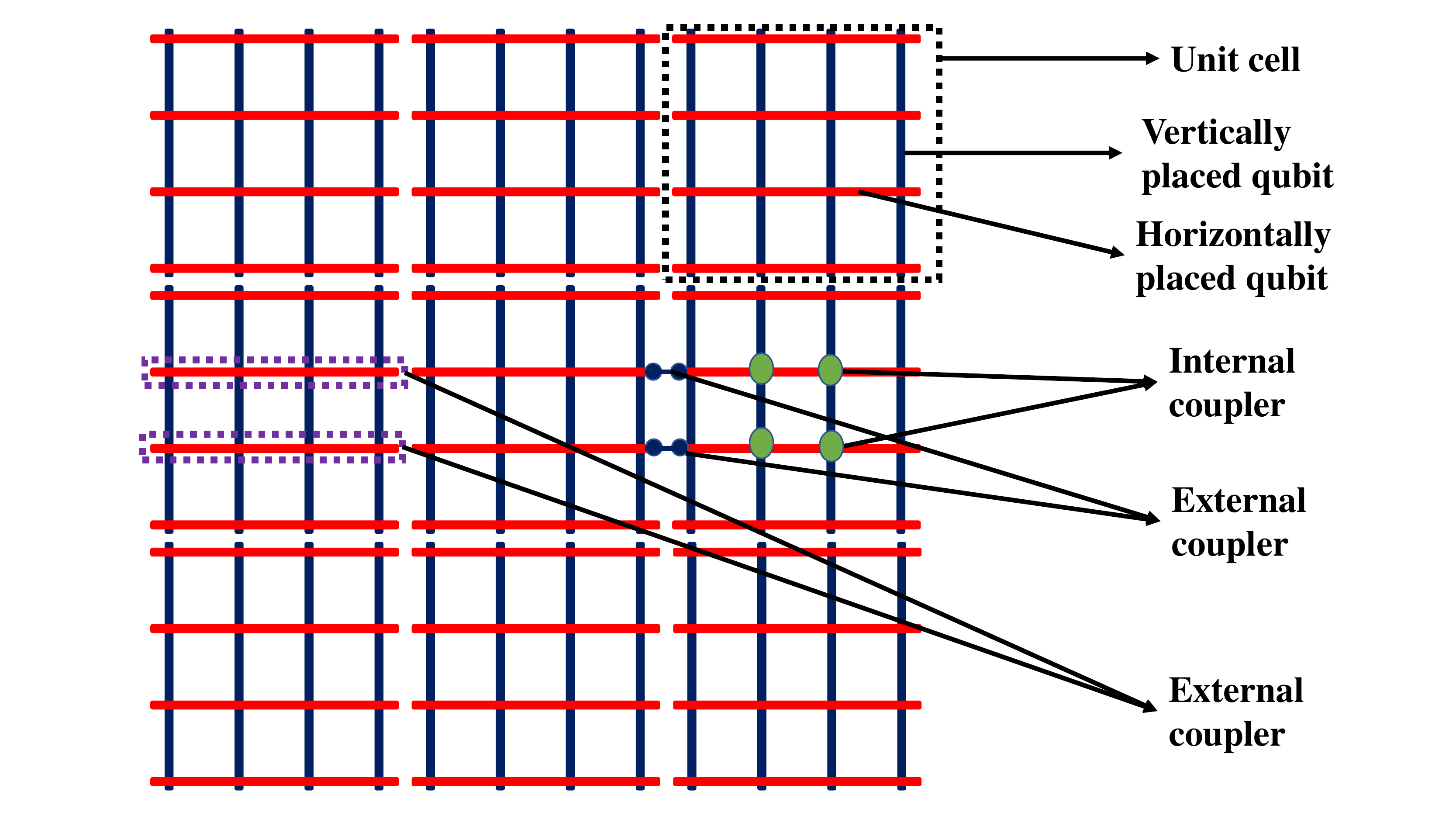}
\caption{Architecture of D-Wave's QPU layout}
\label{qpu_layout_label}
\end{figure*}

Programming into the QPU mainly consisting of initializing and loading the bias and weight associated with qubits. The biases and weights of the spin variables are mapped into the physical qubits of the quantum processor. The qubits are physically arranged in the form of a unit cell, with each unit cell consisting of 8 qubits. Figure \ref{qpu_layout_label} shows a $3 \times 3$ layout of unit cells consisting of a total of 72 qubits.
Out of the 8 qubits, 4 qubits are placed horizontally and 4 are placed vertically. Each vertical qubit is fully connected to the four horizontal qubits and vice-versa. The qubits in a unit cell are connected using internal couplers (shown in green dots). External couplers can be used to connect qubits between two different unit cells and also between vertically placed qubits in the same unit cell. For example, the blue dots connected by a line shows the interconnection between two horizontally placed qubits in two different unit cells and the dotted purple line represents the connection between two vertically placed qubits in the same unit cell. This type of layout is also called a Chimera graph. 

\begin{figure}[h]
\centering
\includegraphics[trim=1cm 0cm 0cm 0cm, scale=0.27]{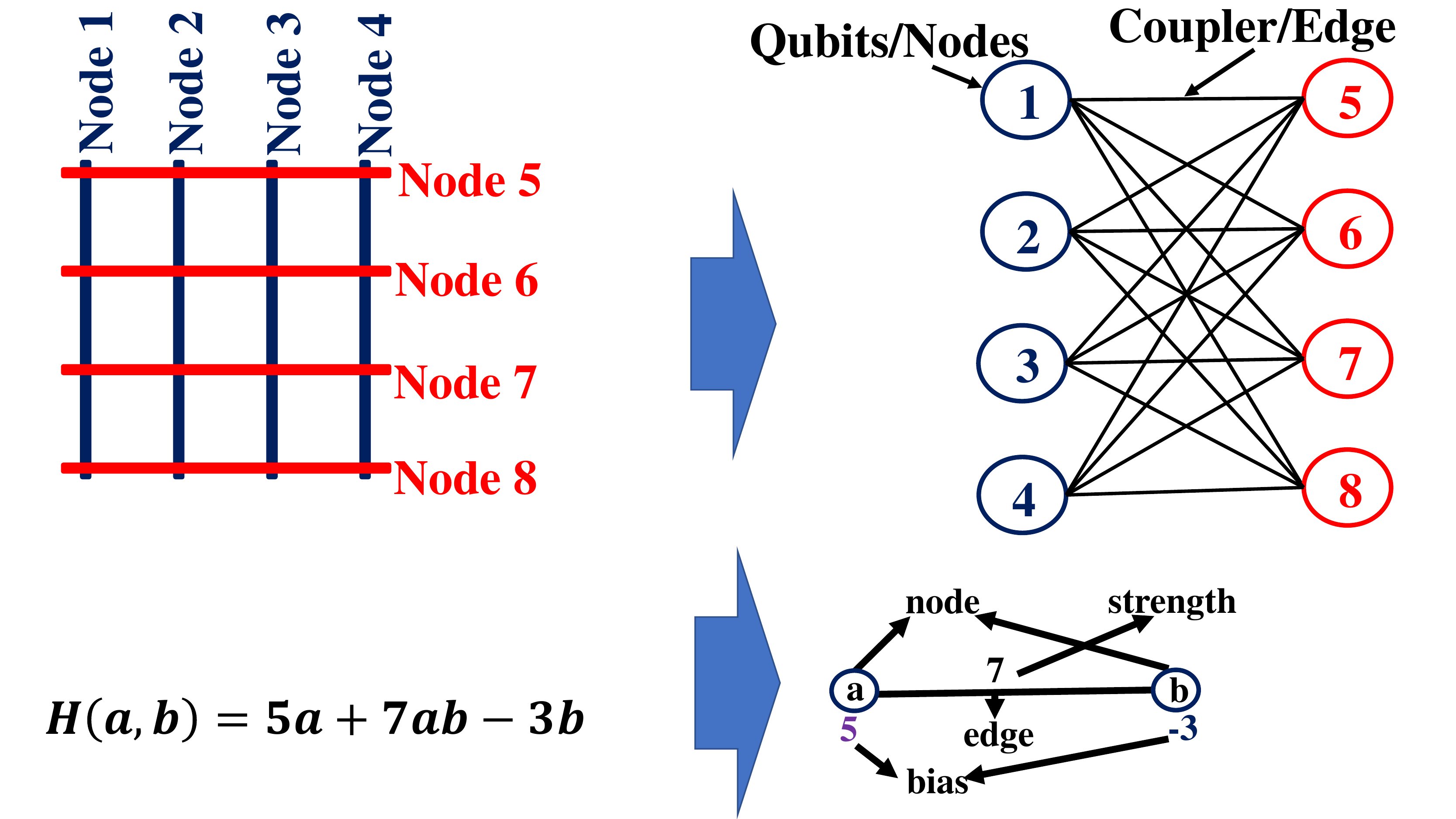}
\caption{Graphical representation of a two-variable objective function}
\label{objective_function_label}
\end{figure}

The qubits of the D-Wave's QPU can be represented as nodes and the edges connecting the nodes as couplers. Figure \ref{objective_function_label} shows a unit cell with 8 qubits. The four qubits shown in blue vertical lines represent nodes 1-4 and the four red horizontal lines represent qubits 5-8. A transformation of the layout to its equivalent graph structure resembles a neural network with two fully connected layers, with four neurons at each layer. Such a structure can be used to represent an objective function in machine learning. For example, if we consider a two-variable objective function, $H(a,b)=5a+7ab-3b$, can be mapped into a two-qubit system as shown in Figure \ref{objective_function_label} (below).

In the objective function, we have two variable and the constant terms associated with the two variables are the bias term that is 5 and 3 in this case. The constant term in the quadratic expression of the two variables, that is 7 in this case is the strength of the coupling between the two variables. The structure of the D-Wave system's QPU supports objective functions with a maximum degree of 2, or in other words, quadratic expression. Alongside Ising, QPU also supports QUBO problems known as Quadratic Unconstrained Binary Optimization (QUBO). The objective function can be mathematically expressed as 
$\underset{x \in (0,1)^{n}}{min} x^{T}Qx $, where $x$ is a binary vector and $Q$ is an upper triangular matrix of real weights, and the objective is to minimize the above expression. 

An important aspect of programming into D-Wave's QPU, is the identification of a representation of the cost function dependency graph within the constraints of the Chimera hardware. A number of techniques have been developed for this graph embedding procedure \cite{037,038}. Embedding increases the number of resources required to represent the original problem but ensures that the correct logical relationships are encoded. Specialized embeddings can be developed for constrained input graph structures, such as the bipartite graph that is frequently used in machine learning \cite{039}. After programming the biases and weights into the chimera graph structure for the given problem instance and initializing other hyperparameters, the problem is sent to the quantum hardware for computation.
\subsubsection{Resample} The solution returned by the quantum annealing process may not end up in the ground state because of the influences of external energy sources \cite{035}. Hence, in practice, the probability of the system staying in a low energy state might be low for certain problems. This can be mitigated by taking several such close-to-optimal solutions and generate a distribution of the possible solutions. The distribution of the possible solutions determines the frequency with which a specific solution is observed.  
\subsection{Quantum Computation}
After the qubits are loaded with biases and weights, the system undergoes quantum annealing to find solutions to the problem instance. This process is distinguished into two parts: (i) Quantum Annealing and (ii) Readout. After the process of quantum annealing, the initial ground energy states evolve to their optimal ground energy states and the biases and weights change to their optimal configurations. After the transition ends, the qubits have classical spin values as governed by the objective function. After annealing, a readout is performed to retrieve the classical spin values and perform a scaling operation to convert the classical spin to real-valued weights. The total time for performing an anneal operation can be estimated by summing up the programming time, annealing time, and readout times. Hence, resampling the solution greatly reduces the total computation time as only the readout phase has to be executed instead of performing the annealing operation several times.

With the background on quantum annealing and the D-Wave's quantum annealing processor, we will now explore the application of quantum annealing to machine learning classification tasks. These theoretical sections on quantum annealing background will help us analyze and understand the different ways in which quantum annealing can be used to improve the efficiency of classification algorithms.

\section{Quantum Annealing for Machine Learning Classification}
\label{works_qa}
In recent years, researchers have been interested in applying quantum annealing to improve various aspects of training a machine learning classifier \cite{003,015,016,018}. These works vary from each other in the ways they have used quantum annealing for training, the physical D-Wave computer on which the QA algorithm is implemented, the dataset, and the application domain. Table \ref{dwave_ver} shows the different versions of D-Wave systems along with attributes like the number of qubits, couplers, and operating temperature \cite{010,011,012,007,018,032}. Recently, a 5000 qubit system has been also launched by D-Wave systems called Advantage. However, because the latest version with 5000 qubits has been launch very recently, it is not in the scope of the discussion of this manuscript. 
\begin{table}[h]
\centering
\caption{Attributes of four versions of the D-Wave systems (Temp is the operating temperature $T_{p}$ is the initialization/programming time, $T_{a}$ is the anneal time and $T_{r}$ is the read out time.)}
\scalebox{0.87}{
\begin{tabular}{|c|c|c|c|c|c|}
\hline 
\textbf{Attributes} & \textbf{One} & \textbf{Two} & \textbf{2X} & \textbf{2000Q} & \textbf{Advantage} \\ 
\hline 
\textbf{Qubits} & 128 & 512 & 1152 & 2048 & 5640 \\ 
\hline 
\textbf{Couplers} & 352 & 1472 & 3360 & 6016 & 40,184 \\ 
\hline 
\textbf{Temp} (K) & NA & 0.02 & 0.015 & 0.015 & 0.015 \\ 
\hline 
\textbf{$T_{p}$} & 270 ms & 36 ms & 10 ms  & NA & NA\\ 
\hline 
\textbf{$T_{a}$} & 1 ms & 20 $\mu$s & 20 $\mu$s & 20 $\mu$s  & 20 $\mu$s\\ 
\hline 
\textbf{$T_{r}$} & 1.5 ms & 0.13 ms & 120 $\mu$s & NA & NA\\ 
\hline 
\end{tabular}  
\label{dwave_ver}
}
\end{table} 


Machine learning tasks using quantum annealing are performed in a hybrid classical and quantum system. The various aspects of training which need to be optimized can be offloaded to the quantum processing unit for speed and accuracy. Researchers have utilized the power of quantum annealing in various aspects of training a machine learning classifier and validated the results on a wide range of dataset suitable for various real-time applications. In this section, we will present and analyze existing works in this area by categorizing the works based on the applications for which quantum annealing was intended. Table \ref{related_work} lists the articles that have reported results on performing machine learning classification tasks using D-Wave's quantum annealer based on recent research works. The works are tabulated by categorizing them based on the intended application, the version of the D-Wave on which the problem was solved, three aspects of training which include the aspect of the training in which quantum annealing was used for and its classical counterpart with which QA was compared. Finally, the table also consists of aspects of the dataset such as source, the size of training, and test data set. Listing these aspects in a tabular form will help us to properly organize our discussion of these works.

Apart from the works discussed here, there are several works that have used quantum annealing for machine learning tasks. For example, recently, quantum annealing was also applied for training a restricted Boltzmann machine for applications in cybersecurity \cite{041}. Apart from the machine learning classification tasks performed on physical D-Wave's quantum computer, several other works have also explored the idea of using quantum annealing for optimization by simulation of the quantum phenomenon instead of the actual hardware. Some of the examples of such works are in the area of lung cancer detection \cite{026}, multi-class classification \cite{027}, training in presence of label noise \cite{028}, natural language processing, seizure prediction, and linear separability testing \cite{029}. However, in this paper, our focus is to discuss and analyze the use of quantum annealing for training machine learning classifiers using domain-specific dataset. Hence, we will be focusing only on those works that have emphasized real-world applications using domain-specific dataset and have used a physical quantum annealer.

\begin{table*}
\centering
\caption{Related work on machine learning classification using quantum annealing for optimizing training. Application refers to the broader on which the use of quantum annealing was validated on. Quantum annealing application summarizes the primary use of quantum annealing in the training procedure and classical comparison refers to the alternative classical training approach that is used for comparison. DCNN=Deep Convolutional Neural Network, DNN=Deep Neural Network, SA=Simulated Annealing, QSA=Quantum Simulated Annealing, MLR=Multivariate Linear Regression, SVM=Support Vector Machine, CD=Contrastive Divergence.}
\scalebox{.87}{
\begin{tabular}{|c|c|c|c|c|c|c|c|}

\hline
 &  &  & \multicolumn{2}{c|}{ \textbf{Training}} & \multicolumn{3}{c|}{ \textbf{Dataset}}\\
\cline{4-8}
 &  & &  & & &  &  \\
 &  &  \textbf{D-Wave} &  \textbf{Quantum Annealing} & \textbf{Classical} &  \textbf{Source} &  \textbf{Train} &  \textbf{Test}\\
 \textbf{Work }&  \textbf{Application} &  \textbf{Version}  &  \textbf{Application} &  \textbf{Comparison} & & \textbf{Size}  & \textbf{Size}\\
\hline
Adachi et al. \cite{003} & Image  & D-Wave 2 & Estimation of & Contrastive Divergence & MNIST & 60,000 &10,000 \\
& Recognition & & model expectation & based training & & &\\
\hline
Nguyen et al. \cite{014} & Image & D-Wave 2X  & Generating sparse & Classical matching pursuit & MNIST & 50000 & 5000 \\
&  Recognition &  & Representation & SVM, Alex Net like DCNN & & &\\
\hline
Boyda et al. \cite{015} & Remote Sensing & D-Wave 2X & Selecting optimal & Simulated Annealing & NAIP & 24,000 & 6,000 \\
& Imagery & & voting subset & & & & \\
\hline
Mott et al. \cite{016} & Particle Physics & D-Wave 2X  & Selecting optimal set & DNN, XGBoost & Simulated & 20 x  & NA \\
&  & & of weak classifiers & & & 20000 & \\
\hline
Li et al. \cite{017} & Computational & D-Wave 2X & Estimation of & SA, SQA, MLR, & gcPBM & 1500 & 166 \\
& Biology & & feature weights & Lasso, XGBoost & HT-SELEX & 4500 & 500 \\
\hline
Willsch et al. \cite{018} & Computational & 2000Q & Quantum SVM & Classical SVM  &  \cite{019} & 4352 & 484 \\
& Biology & &  &  & & &  \\
\hline
Dixit et al. \cite{020} & Image & 2000Q & Estimation of  & Contrastive Divergence & BAS & 300 & 200 \\
& Recognition &  & model expectation & based training & & & \\
\hline 
Caldeira et al. \cite{024} & Image & 2000Q & Estimation of & CD, SA, & Galaxy Zoo & NA & NA\\
& Recognition & & model expectation & Gradient Boosted DT & & & \\   
\hline
Liu et al. \cite{032} & Image & D-Wave 2X & Estimation of & High Performance  & MNIST & NA & NA \\
& Recognition &  & model expectation & Computing & & &\\
& Particle Physics & & & Neuromorphic Computing  & \cite{033} & NA & NA\\


  \hline
\end{tabular}
}
\label{related_work}
\end{table*}

\subsection{\textbf{ML Applications Using D-Wave's Quantum Annealing}}
\begin{figure*}[h]
\centering
\includegraphics[trim=1cm 3.5cm 0cm 0cm, scale=0.5]{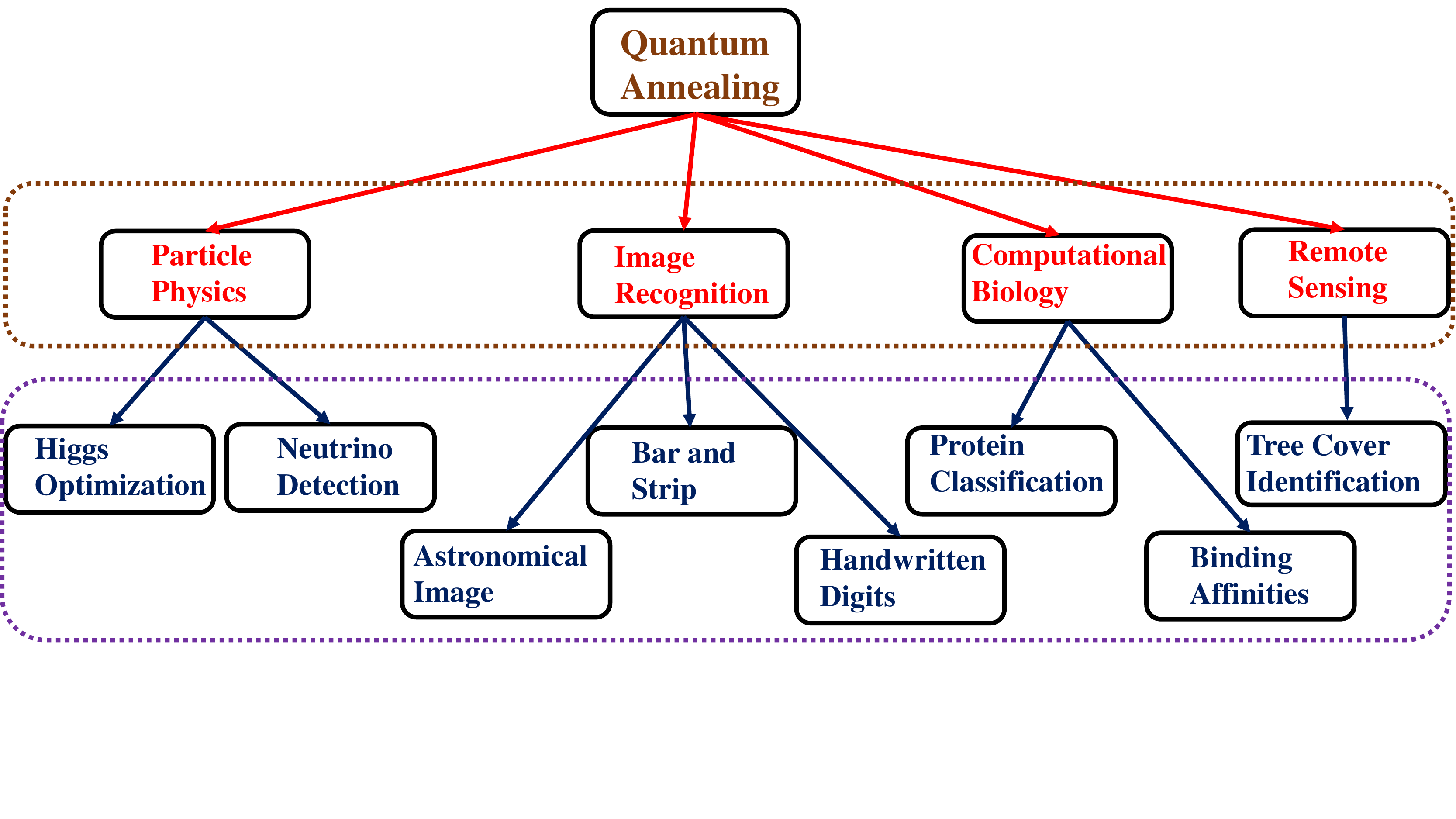}
\caption{Representative works in some of the domains where quantum annealing was applied in machine learning}
\label{concept_application_label}
\end{figure*}

\begin{figure*}[h]
\centering
\includegraphics[trim=1cm 0cm 0cm 0cm, scale=0.5]{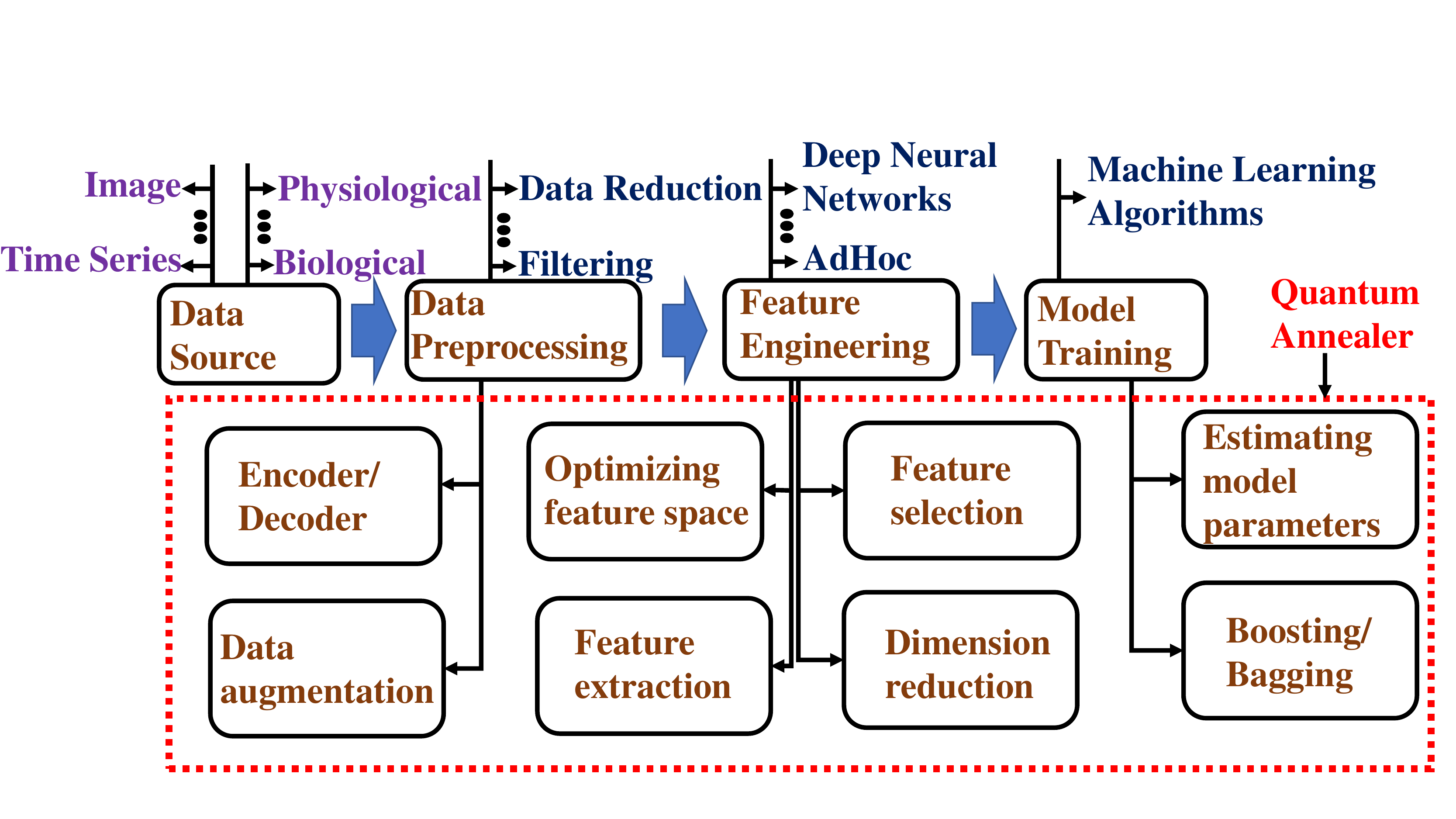}
\caption{Example of a generic machine learning pipeline illustrating some of the phases that can be optimized using a quantum annealer.}
\label{hybrid_overview_label}
\end{figure*}

In the current research landscape, researchers have used quantum annealing for real-world application using domain-specific dataset in four broad areas: (i) Image Recognition (ii) Remote Sensing Imagery, (iii) Particle Physics, and (iv) Computational Biology (Figure \ref{concept_application_label}). The representative works on these domains will be discussed in detail in this section. The works discussed here have adopted a hybrid classical-quantum system for optimizing machine learning classifiers. Based on the works discussed here, an overview of such a hybrid system can be illustrated with the help of Figure \ref{hybrid_overview_label}.

\subsubsection{Image Recognition}
Work done in \cite{003,014,020, 024} has investigated the effectiveness of using quantum annealing in image recognition tasks. Qualitatively, \cite{003,020} have used QA in a similar way but the implementation varied with respect to the D-Wave version used and the dataset used for validation. For \cite{003}, the MNIST data set was used and the classifier's task was to recognize handwritten digit (0-9), whereas in \cite{020}, a simple bar and strip dataset was used. Further, the training size used in \cite{003} is about 60,000 samples and that of \cite{020} had about 300 training samples. On the other spectrum, \cite{003} and \cite{014} used two different approaches in using QA but the validation was done on the same dataset (MNIST) with 50000 training samples.  A similar idea of RBM training using QA to classify the shape of galaxies is used in \cite{024}. 

In \cite{003}, authors used quantum annealing to determine the model expectation of a Deep Belief Network (DBN) which is a Deep Neural Network with stacked Restricted Boltzmann Machine \cite{021}. The model expectation of a Deep Belief Network is typically performed using Contrastive Divergence. The experiment was conducted on the MNIST dataset which is a benchmark dataset for image recognition \cite{022}. The dataset consists of handwritten digits between 0 and 9 with dimension $28 \times 28$. To compare the quantum and the classical methods, two DNNs of the same size were implemented. Each DNN had 32 input nodes, two hidden layers followed by 10 output layers. The training was divided into generative (quantum annealing is used) and discriminative (completely classical). For the quantum model, quantum annealing was used to estimate model expectation in the generative phase and for the classical model, Contrastive Divergence was used. The generative phase calculates updates to the weights and biases and the discriminative phase uses backpropagation for fine-tuning the biases and weights. Results showed that the quantum model could achieve a specific level of accuracy with fewer iterations of generative and discriminative training. A quantitative evaluation showed that the quantum model outperforms the classical model in terms of accuracy with only 20 and 100 generative and discriminative iterations respectively. The classical model took about 50 generative and 800 discriminative iterations to reach the same level of accuracy. 

A similar approach was used in \cite{020}, however, the idea was tested on BAS (bar and strip database) and on D-Wave 2000Q. However, instead of a DBN, the authors used a single RBM unit with a visible layer and hidden layer. The article focused on the comparison of the quantum implementation of RBM and classical implementation of RBM on the D-Wave computer. The implementation of the RBM in the quantum model had 64 visible units and 64 hidden units with all nodes of the visible layer fully connected to all nodes of the hidden layer. However, due to faulty qubits, some connections were non-functional as was the case in \cite{003}. The structure of the RBM in the classical model was also the same. The only difference between the two models is that the quantum model estimated model expectation of the RBM using quantum annealing as opposed to CD-based training in the classical model. Both the models were validated on a BAS dataset containing 500 records consisting of 64 bit variables each. The first 62 bits represent the image and the last two bits represent its label (Bar or Strips). Results showed that both the approaches gave comparable classification performance, however, higher fluctuations in the classification accuracy were observed with CD-based training.

In \cite{014}, the implementation of recognizing handwritten digits was implemented on D-Wave 2X. This work followed a different approach to using quantum annealing for training. In this work, the authors used QA to generate a sparse representation of the images. First, the images were downsampled and were fed into a fully connected graph of 47 logical qubits. The objective of the quantum annealing procedure, in this case, was to obtain an optimal sparse representation of the image that can most accurately represent the original image. The method was compared with the classical approach in two ways: (i) First, the sparse representation generated by QA was fed to a softmax classifier and an SVM classifier and the result was compared with the sparse representation generated by Alex Net like DCNN. (ii) An equivalent classical sparse representation algorithm known as the classical matching pursuit was used to find the optimal sparse representation of the images. The resulting sparse representation from QA based approach and classical approach were fed into a Multi-Layer Perceptron (MLP) for classification. In both cases, superior performance of the quantum-based approach was observed. Another comparison was done with state-of-the-art RESNET (with a large number of neurons and parameters) which showed slightly better performance than the quantum approach. However, the authors found that the accuracy of RESNET decreased as the number of training samples was reduced. 

In \cite{024}, the idea of using QA to estimate model expectation of an RBM machine in classifying the morphology of galaxies. The model was implemented on a D-Wave 2000Q computer with ~2000 qubits. The quantum approach was compared to Contrastive Divergence based training, Simulated Annealing and gradient boosted decision trees. Galaxy Zoo dataset was used for this experiment. The images of the galaxies were first compressed using principle component analysis to represent the image in the physical hardware. Experimental results showed that quantum annealing could perform better than classical algorithms if the dataset is small and with a limited number of training iterations. Overall, the authors did not find any algorithmic advantage of using QA over the classical approach for this particular dataset. In \cite{032}, quantum annealing was used to train a Limited Boltzmann Machine, which is a variation of Restricted Boltzmann Machine to recognize handwritten digits. 

\subsubsection{Remote Sensing Imagery}
In \cite{015}, researchers have investigated the application of quantum annealing in a remote sensing imagery problem. The problem objective was to classify image data as either covered with trees or not covered with trees. In this problem, quantum annealing was used to select an optimal set of voting classifiers that could most accurately classify a given segment as covered with trees or not. The idea was to build around a modified implementation of QBoost \cite{023} which is a quantum version of boosting algorithm. 537 image tiles were taken from the National Agriculture Imagery Program (NAIP). The selected image tiles contained 30000 8x8 labeled pixel squares. 24000 samples were used for training and 6000 were used for validation and testing. 112 features were extracted from the image tiles which resulted in 224 weak classifiers. After performing some initial preprocessing, like discarding classifiers that did not perform better than random guessing, 108 weak classifiers remained. Quantum annealing was used to select the optimal set of weak classifiers. This method was compared with simulated annealing and results showed that both these methods performed comparably in terms of optimal solution and rate of convergence.

\subsubsection{Computational Biology} 
The application of quantum annealing in solving computational biology problems has been investigated in \cite{017,018}. In \cite{018}, the objective was to classify a particular protein on its ability to bind with a particular DNA sequence. Whereas in \cite{017}, the objective was to classify the strength of the binding of a protein and a DNA sequence based on a pre-determined threshold.

In \cite{018}, the authors implemented a quantum version of Support Vector Machine by representing the objective function of the SVM as QUBO formulation on D-Wave 2000Q machine. Subsequently, quantum annealing was used to estimate the weights of the objective function. This method was compared with the classical implementation of SVM. The modeling was performed on an experimental dataset \cite{019}, with 4352 training samples and 454 test samples. Results showed that the combination of multiple solutions obtained from D-Wave quantum annealer proved to be superior to the single solution obtained from classical SVM. The authors also observed that the combination of the multiple solutions returned by quantum annealing could possibly generalize better to unseen data than the classical SVM. Based on the findings, the authors have concluded that the quantum version of SVM could find useful practical applications for hard classification problems where sufficient training data is not available. 

In \cite{017}, QA was used to estimate the weights of an objective function of the form $F(\vec{w}) = R(\vec{w})+\Omega(\vec{w})$, where $R$, $\Omega$, and $\vec{w}$ are the loss, regularization and feature weights respectively. The quantum approach was compared with Simulated Annealing (SA), Simulated Quantum Annealing (SQA), Multiple Linear Regression (MLR), Lasso, XGBoost. The objective of the experiment was to compare the performance of the quantum and classical methods in classifying TF-DNA (Transcription Factor-DNA) binding. The experiment was conducted on two types of dataset, one is the gcPBM (genomic-context protein binding microarray) with 1500 training and about 166 test samples. The second is the HT-SELEX (high-throughput systematic evolution of ligands by exponential enrichment) with 4500 train and 500 test sizes. D-Wave 2X was used for this experiment with ~1000 qubits. Experimental results on the two dataset showed that the QA-based approach could outperform the classical technique by a slight margin if the training data is limited in size. Another interesting finding reported by the research is the ability of a quantum-based solution to learn relevant biological patterns. 

\subsubsection{Particle Physics}
Quantum annealing has also been applied to a particle physics problem in \cite{016}. The primary objective of the experiment was to evaluate the efficiency of quantum annealing in classifying Higgs signal from background noise. The quantum method was compared with the classical Deep Neural network and XGBoost framework. QA was used in the selection of the optimal number of weak classifiers from a total of 36 weak classifiers. The weak classifiers were constructed from the distribution of kinematic variables. The experiment was conducted in D-Wave 2X with ~1000 qubits. A simulated dataset was used for the experiment with a 20 x 20000 training dataset. Experimental results showed that the QA-based approach was not very efficient in finding the true minimum. The reason for this has been attributed to the noise in the physical system which is exacerbated by the sparse connectivity among qubits. Further investigation with dataset of varying sizes showed the clear advantage of using the quantum approach over the classical approach for optimization. However, the advantage diminishes as the dataset grows larger. Another work in \cite{032} explored the application of quantum annealing in neutrino detection dataset.

\section{Advantages of using QA for Training ML Models}
\label{advantage_qa}
Analyzing the experimental results on the existing works on using quantum annealing for training, the following possible advantages of using QA for machine learning classification can be inferred.

\subsection{\textbf{Training under limited data}} Quantum annealing could be a better choice for training ML classifiers in the presence of limited training data. Results from \cite{014,024,018,017,016} consistently reported higher performance of quantum annealing based approach when the dataset is small. For smaller dataset, the performance of QA was reported to be better than that of classical approaches. However, classical approaches outperformed quantum-based training when the size of the data is increased. The poor performance of the quantum approach was attributed to the hardware constraint of the quantum hardware. Further research on different dataset is needed to validate the effectiveness of using quantum annealing for training models with limited data. 

\subsection{\textbf{Dimension Reduction}} Feature selection is an important subroutine for optimizing the training and performance of machine learning models and experimental results indicate the potential of QA over the classical approach for dimension reduction \cite{014,040}. Results reported in \cite{014} confirmed the superior performance of the sparse representation generated by QA over the classical approach. Further, experimental results based on the application of QA to remote sensing imagery \cite{015}, reported that the QA-based approach of selecting an optimal voting subset from feature variables outperformed simulated annealing based approach. 

\subsection{\textbf{Generalization to Test Data:}} Machine learning classifiers based on quantum computation are hypothesized to be able to generalize better to unseen data than purely classical machine learning classifiers \cite{025,018,015}. Results reported in \cite{018}, showed that the solutions returned by the quantum version of SVM were able to better generalize to new data as opposed to the classical version of SVM. Also, the results reported in the remote sensing imagery problem \cite{015}, was also consistent with the strong generalization performance of the the QA-based classification approach. However, further research needs to be done to validate the generalization performance on different data sets and also for larger dataset.

\subsection{\textbf{Multiple Solutions}} Quantum annealing can also possibly yield many counter-intuitive solutions to a given optimization problem. In the readout phase (Figure \ref{qpu_overview_label}), qubits stay in a superposition state. The real-valued weights and biases are obtained by scaling the classical spin values of the qubit. Hence, every time the readout operation is performed, the hardware returns slightly different solutions. This is validated in the experiment performed in \cite{018} showed that the quantum version of SVM was able to generate several solutions as compared to the single solution obtained by the classical SVM which are also close to optimal solutions. Further experiments performed in the application of quantum annealing for signal isolation \cite{016} also highlight the superior performance of quantum annealing in generating diverse solutions. 

\subsection{\textbf{Reduced Training Time}} Quantum computing techniques owing to their superposition and entanglement phenomenon can evaluate $2^{N}$ states at once along with speed up due to quantum tunneling. Because of this, it is hypothesized that quantum computing strategies can achieve significant speed-ups in machine learning applications. This reduction in training time will also make the model learning more effective in production when it has to update its weights in real-time in presence of new data. Initial experimental results concerning speed-ups have shown promise of the theoretical results derived so far.
In \cite{003,024}, researchers noticed superior performance of the QA-based training when the training iteration was limited. This implies that given similar training conditions for quantum and classical approaches, it is expected that the quantum-based training approach will converge to the solution with fewer iterations than the classical approach. However, further investigations with rigorous experimental proof are required to experimentally validate the effect of quantum speed-up in reducing training time.

\section{Conclusions and Discussions}
\label{conc}
In this article, we have explored a novel emerging computing paradigm known as quantum annealing and its application in the context of machine learning classification. We have presented a detailed survey of the existing efforts in the direction of applying quantum annealing for improving various aspects of training a machine learning classifier. Our survey has been limited to those demonstrations that emphasized real-world applications. We have discussed the background of quantum annealing and its implementation in D-Wave's quantum computer. We have categorized and analyzed the existing works that have used D-Wave's quantum computer for machine learning classification based on the application domain. Based on the experimental results and analysis, we have also listed several aspects of machine learning classification that can be better handled with a hybrid classical-quantum system. Quantum annealing has immense potential in advancing state-of-the-art machine learning classifiers through improvement in speed and performance. However, several challenges need to be overcome before quantum annealing can be widely used as an alternative for classical computation for classification tasks. This includes but is not limited to an increase in the number of qubits, the connections among the qubits, etc. Despite the limitations faced currently, the current trend indicates a bright possibility of successfully incorporating quantum computation techniques in everyday real-world applications in the near future.


\section{Compliance with Ethical Standards}
\textbf{Conflict of Interest}: The authors declare that they have no conflict of interest.\\
\textbf{Ethical approval}: This article does not contain any studies with human participants or animals performed by any of the authors.

\bibliographystyle{spmpsci} 
\bibliography{references}

\end{document}